# Does Non-Genetic Heterogeneity Facilitate the Development of Genetic Drug Resistance?

Kevin S. Farquhar[1], Samira Rasouli Koohi[2], and Daniel A. Charlebois[2,3,*]
[1]Precision for Medicine, Houston, TX, 77054, USA
[2]Department of Physics, University of Alberta, Edmonton, AB, T6G-2E1, Canada
[3]Department of Biological Sciences, University of Alberta, Edmonton, AB, T6G-2E9, Canada
*Corresponding Author: dcharleb@ualberta.ca

**Abstract**

Non-genetic forms of antimicrobial drug resistance can result from cell-to-cell variability that is not encoded in the genetic material. Data from recent studies also suggest that non-genetic mechanisms can facilitate the development of genetic drug resistance. In this Perspective article, we speculate on how the interplay between non-genetic and genetic mechanisms may affect microbial adaptation and evolution during drug treatment. We argue that cellular heterogeneity arising from fluctuations in gene expression, epigenetic modifications, as well as genetic changes contributes to drug resistance at different timescales, and that the interplay between these mechanisms may influence the evolutionary dynamics of pathogen resistance. Accordingly, developing a better understanding of non-genetic mechanisms in drug resistance and how they interact with genetic mechanisms will enhance our ability to combat antimicrobial resistance.

**Introduction**

Antimicrobial (drug) resistance is defined as a heritable decline in the drug sensitivity of a microbe and is a well-known consequence of evolution by natural selection [1]. During this process, microorganisms that are the least sensitive to the drug increase in frequency and pass along genetic material that confers resistance to other cells.

The development of drug resistance is traditionally thought to arise from heritable acquired mutations, or from the horizontal transfer of genetic material from a non-parental donor through conjugation, transformation, or transduction [1]. This horizontal gene transfer is a major source of genetic diversity in prokaryotes [2], and it is not surprising that genetic mechanisms have been largely been thought of as the sole driver of antimicrobial resistance. We believe that this may be an oversimplification of a more complex reality. Notably, in some cases the low rate of beneficial mutations [3,4] seems incompatible with the rapid onset of drug resistance [5,6].

Recent studies have demonstrated that drug resistance can arise in bacterial [7], fungal [8,9], and mammalian cells [10] from non-genetic mechanisms that do not involve changes in DNA sequences. Instead, the resistance arises from a diversification of phenotypes within populations of genetically identical cells in the same environments [11]. This non-genetic diversification can in turn lead to the emergence of a phenotypically heterogeneous cell population in which some cells have a better ability to withstand drug exposure. Non-genetic heterogeneity within microbial populations can arise from epigenetic mechanisms (i.e., DNA methylation, histone modifications, and chromatin structure) as well as from variability in the expression of a gene

[12]. Fluctuations in gene expression are referred to as gene expression "noise" [13], though biological noise is inherent in every biochemical process including the epigenetic modification of DNA [12]. Gene expression noise can lead to the evasion of successful drug therapy through the transient expression of resistance genes [5,14]. Non-genetic phenomenon arising from distinct molecular processes can lead to heritability, in some cases by induction from drugs. Therefore, an understanding of non-genetic heritability and heterogeneity is necessary for tackling drug resistance.

Phenotypic heterogeneity has been predicted to alter the evolutionary dynamics of populations under stress by transforming the fitness landscape [15-17]. It has been suggested that non-genetic heterogeneity can accelerate the rate of adaptive evolution by rapidly generating subpopulations with novel phenotypic traits in populations facing extreme environmental challenges [16]. These studies raise the fundamental question of how non-genetic mechanisms fit into the theory of evolution. Here, we propose that non-genetic and genetic mechanisms spanning different timescales interact to facilitate the survival and evolution of drug-resistant microbes.

**Whence does non-genetic heterogeneity in drug resistance arise?**

Non-genetic heterogeneity in the abundance of gene expression products results from inherently noisy biochemical reactions [13], as well as several other factors including environmental conditions [18], and is modulated by gene-specific and genome-wide processes [19]. Here, we will refer to the variability inherent in the biochemical process of transcribing a messenger RNA (mRNA) molecule from a gene, and the subsequent translation of a protein from mRNA, as transcriptional variability [20]. The transcriptional variability of a drug resistance gene can be modulated by the "architecture" of the gene regulatory network (i.e., how the genes in the network biochemically regulate one another) in which it is imbedded [21]. For instance, feedforward and positive feedback regulation in fungi have been found to increase the timescale of non-genetic phenotypes, such that surviving non-genetically drug-resistant cells can continue to divide and enhance the reproductive fitness of the population [8,22]. Mammalian cell lines with a high-noise positive feedback drug resistance gene network have been observed to facilitate adaptation under high-drug concentrations, compared to mammalian cell lines with a low-noise negative feedback drug resistance gene network controlling the same resistance gene [10]. It is important to note that the non-genetic variability in the gene expression process is, in part, encoded in the promoter DNA sequence of the gene [23], as is the architecture of genetic networks.

Epigenetic heterogeneity is another important contributor to non-genetic heterogeneity [12]. DNA methylation is an epigenetic mechanism that normally silences gene expression [24]. There is a widespread heterogeneity associated with DNA methylation between genomes, which is thought to increase gene expression variability [25]. Additionally, stress from toxins or drugs also influence DNA methylation patterns [26]. Histone modifications, another epigenetic mechanism, are small chemical moieties that are covalently attached to subsections of histone proteins, which can activate or repress gene expression. This epigenetic form of gene regulation is achieved by the recruitment of transcription factors or by influencing accessibility to DNA [27], which can modulate gene expression variability through transcriptional "bursting" [28]. Nucleosome positioning modulated by chromatin remodeling activity can also act as a driver for gene expression variability [29]. At higher levels of chromatin organization, a tightly packed form of DNA called heterochromatin can increase gene expression variability because the spatial expansion or reduction of regions in the heterochromatin state can itself vary randomly [30].

In summary, phenotypic variability impacting drug resistance has been demonstrated to arise from multiple scales of transcriptional variability, histone modification, chromatin remodeling, and other epigenetic modifications of DNA, as well as genetic variability.

**Heritable timescales for drug resistance evolution**

Gene expression fluctuations can be classified as "intrinsic" or "extrinsic" noise [21,31]. Intrinsic noise, resulting from the inherent randomness in the biochemical processes of transcription and translation [31], is not a useful substrate for natural selection, as these timescales are too short (e.g., the autocorrelation time for intrinsic noise in *E. coli* bacteria is ≤10 minutes, which is shorter than its cell cycle time [32]) to impact heritable drug resistance in microbes (Fig. 1). On the other hand, extrinsic noise, due to the fluctuations in the amounts or states of other cellular components that lead indirectly to variability in gene expression [31], may lead to heritable drug resistance [14], as these fluctuations persist over a longer timescale (e.g., the autocorrelation time for extrinsic noise in *E. coli* is ~40 minutes, which is similar to its cell cycle time [32]). Furthermore, when fluctuations in gene expression are modulated by a gene regulatory network with the appropriate architecture, the increased duration of the fluctuations can facilitate adaptation during drug treatment [8,22]. For instance, the fluctuations timescale for drug resistance genes regulated by positive feedback gene networks have been estimated to be 58 hours in Chinese hamster ovary (CHO) cells [10] and 283 hours in the budding yeast *Saccharomyces cerevisiae* [33]. These timescales are much longer than the corresponding cell division time for CHO cells (~18 hours) and budding yeast cells (~2 to 4 hours).

Epigenetic phenomena arising from distinct molecular processes can lead to heritability [6]. Some epigenetic mechanisms are heritable over long timescales, while others disappear within a few generations [34]. DNA methylation is the most heritable epigenetic mechanism, which can also contribute to the heritability of other epigenetic mechanisms such as histone modifications [24]. Inheritance is set by the re-addition of methyl groups on the newly synthesized strand hybridized to hemi-methylated DNA (i.e., when only one of the complementary DNA strands is methylated) by maintenance DNA methyltransferases during cell division [35]. Novel methylation events are introduced by *de novo* DNA methyltransferases; DNA demethylation (which reverses DNA methylation) can subsequently occur through dilution and active demethylation enzymes. In the short-term, DNA methylation patterns can last for at least one cell division due to the requirement of hemi-methylated DNA in the cell cycle. The rate of DNA methylation gain or loss (~4 x $10^{-4}$ per CG pair per generation) also suggests that methylation and demethylation events are unlikely to occur at one site between two subsequent cell divisions [36]. In the longer-term, in the pathogenic fungus *Cryptococcus neoformans*, loss of a *de novo* DNA methyltransferase over 100 million years ago did not prevent the maintenance of DNA methylation patterns because the remaining maintenance DNA methyltransferase enzyme kept the pattern stable over millions of years [37].

The timescales of histone modification heritability are less clear than for DNA methylation. At the shortest timescale, the inheritance of a histone modification has been shown to occur over multiple generations [38]. Heterochromatin stands out as one of the more stable epigenetic processes that is heritable over long epigenetic timescales [39]. Evidence has also been found for the long-term inheritance of some histone modifications but not others in research on engineered gene networks [40]. Many types of organisms have demonstrated inherited forms of histone modifications induced after stress, including pathogenic fungi such as *Candida albicans* [41]. It is thought that histone modifications are not strictly heritable on their own, but require coupling to other epigenetic processes for heritability [42]. More experimental investigations are required to elucidate the stability of histone modifications inheritance over time.

The main insight from these studies is that there does not exist a strict dichotomy between the heritability of non-genetic and genetic mechanisms. Rather, non-genetic states and genetic mutations occupy different positions on a heritability spectrum (Fig. 1). Importantly, both non-genetic and genetic mechanisms can provide the substrate on which natural selection can act. Genetic mechanisms provide substrate with the longest timescales (though not necessarily permanent, as mutations, or even entire genes, can be lost to the cell) and non-genetic mechanisms provide substrate with shorter timescales.

**Interaction between non-genetic and genetic drug resistance mechanisms**

Studies have suggested that non-genetic heterogeneity may have evolved through genetic modifications. For example, promoter mutations can change gene expression noise levels while leaving mean expression levels unchanged [23]. The phenotypic heterogeneity resulting from these genetic changes is considered to be non-genetic once all the cells in the population have acquired the mutation. It has also been recently demonstrated that inherited epigenetic factors contribute to the evolutionary adaptation of gene networks (which are known to modulate gene expression noise [8,22,33]) under sustained selective pressure [43].

Gene expression noise has been proposed to facilitate the genetic evolution of drug resistance [5,14]. There is evidence that elevated transcriptional variability is a selected trait in stress response genes [20]. Non-genetically high expressing budding yeast cells were found to survive drug stress and to sustain the population until more potent drug resistance mutations arose [9]. Non-genetic phenotypic heterogeneity may also increase the net adaptive value of beneficial mutations by generating individuals with exceptionally high trait values at an early stage of adaptation [16]. More specifically, it has been proposed that the phenotypic effects of the mutations that accumulated during the course of microbial drug resistance experiments are contingent on phenotypic heterogeneity. Though the exact mechanism remains unknown, the data suggest that phenotypic heterogeneity may enlarge the set of adaptive mutations that provide resistance above a critical stress level [5,16], and that phenotypic heterogeneity can alleviate the fitness costs of protein expression under a range of stress conditions [16]. The ratio of the physical timescales to biological timescales may determine the degree to which microbial adaptions are non-genetic or genetic through evolutionary trade-offs [44].

Key insights on how epigenetic inheritance interacts with genetic change were revealed during an evolution experiment on budding yeast carrying an auxotrophic selection marker [45]. First, the initial resistance under 5-Fluoroorotic acid (5-FOA) selection occurred most rapidly at the highest levels of epigenetic silencing in the form of heterochromatin, as a result of short-term selection for cells with an inactive URA3 gene. Second, the rate of adaptation from mutations disrupting URA3 activity was observed to be highest at intermediate epigenetic silencing levels, which increased the chance of mutations to arise and provide long-term resistance. Third, the level of mutational supply available outside URA3 that can disrupt sensitivity to 5-FOA was found to be greatest under intermediate levels of chromatin-mediated silencing; this enhanced the heritability of chromatin silencing by increasing the silencing switching rate, which corresponded to enhanced fitness levels of mutated cells. Therefore, heritable chromatin modifications have the capability to improve adaptability in response to stress by increasing the supply of mutations and enhancing heritable silencing through mutations.

Evolution and adaptation to the environment depend on both genetic variation and epigenetic changes (also known as "epimutations"). The mutational supply for epimutations is generally larger than genetic mutations due to the lower rate of genetic mutations per base site per generation [36,46]. Additionally, DNA methylation has the capability to increase mutation rates locally [47]. The primary mechanism is based on the associated biochemical reactions involved after DNA methylation. Importantly, DNA methylation is an intrinsic mediator of genetic mutations, which have the possibility to affect adaptation. Reversing the relationship, single genetic mutations can disrupt DNA methylation patterns [48].

Genetic variability can lead to epigenetic variability, as genetic factors play an important role in epigenetic regulation [18]. It has been hypothesized that when natural selection acts on epigenetic and genetic variation that the adaptive phenotypes arise before genetic changes and the population adapts faster [49]. The interplay between the epigenetic component of accumulating environmental exposure and genetic factors has been proposed as an explanation for the observed discordance between monozygotic twins in terms of disease, such as diabetes [50].

Overall, it has been proposed that non-genetic and genetic mechanisms interact to undermine drug treatment (Fig. 2). Non-genetic mechanisms operate at shorter timescales than genetic mechanisms to produce an acute, reversible response, while genetic mechanisms operate over longer timescales, and in some cases, may assimilate the non-genetically conferred phenotype into a more permanent response. An interesting hypothesis emerges when the cost of drug resistance is considered. That is, if drug treatment fluctuating and of short duration, then resistance may primarily arise through non-genetic mechanisms that minimize the cost of resistance in absence of the drug. Whereas if drug treatment is constant and of long duration, the resistance may be more likely to occur through genetic adaptation.

## Conclusions

How phenotypic heterogeneity promotes evolvability is one of the most important questions in drug resistance research today. The shortage of quantitative modeling and experimental studies investigating the non-genetic variation preceding the genetic changes required for the evolution of complex traits presents an opportunity for researchers. Combining molecular and synthetic biology techniques with microbial evolution experiments could identify the unknown molecular mechanisms by which gene expression noise shape the beneficial effects of mutation. Another important step will be to explore the generality of the role that non-genetic effects play in the specific molecular mechanisms by which different microbes have evolved to resist antimicrobial agents [1]. Finally, the role of non-genetic drug resistance must be established in non-model systems, such as pathogens switching gene expression states to evade host immune responses [6].

When Charles Darwin proposed that heritable variation was the essential ingredient for evolution by natural selection he had no knowledge of the molecular details of the units of heritability. Yet mainstream evolutionary theory has come to focus almost exclusively on genetic inheritance and on genetic mutation as the mechanism for phenotypic variation [6]. However, heritable non-genetic variability is proving to be an important substrate for the development of drug resistance. Furthermore, non-genetic mechanisms are now thought to affect the dynamics of genetic evolution by shaping evolutionary trajectories at the genomic level and facilitating evolutionary rescue during environmental stress [16]. This warrants the attention of drug resistance researchers and evolutionary biologists. Ultimately, collaboration and interdisciplinary research will be required to develop a unified theory of evolution and tackle antimicrobial resistance.

**Author Contributions**

DC conceptualized and supervised the project. DC, KF, and SRK, all contributed to the literature review. DC developed the figures. DC, KF, and SRK wrote the manuscript.

**Acknowledgments**

The authors are grateful to Dr. Michael Manhart, Dr. Mads Kærn, and Dr. Sui Huang for helpful comments on the manuscript. We also thank Harold Flohr and Joshua Guthrie for proofreading.

**Funding**

Precision provided indirect funding support for this paper by way of employment of KF. No additional funding was provided for this paper by Precision or any other commercial entity.

DC is supported in part by funding from the Government of Canada's New Frontiers in Research Fund (NFRFE-2019-01208).

**Figures**

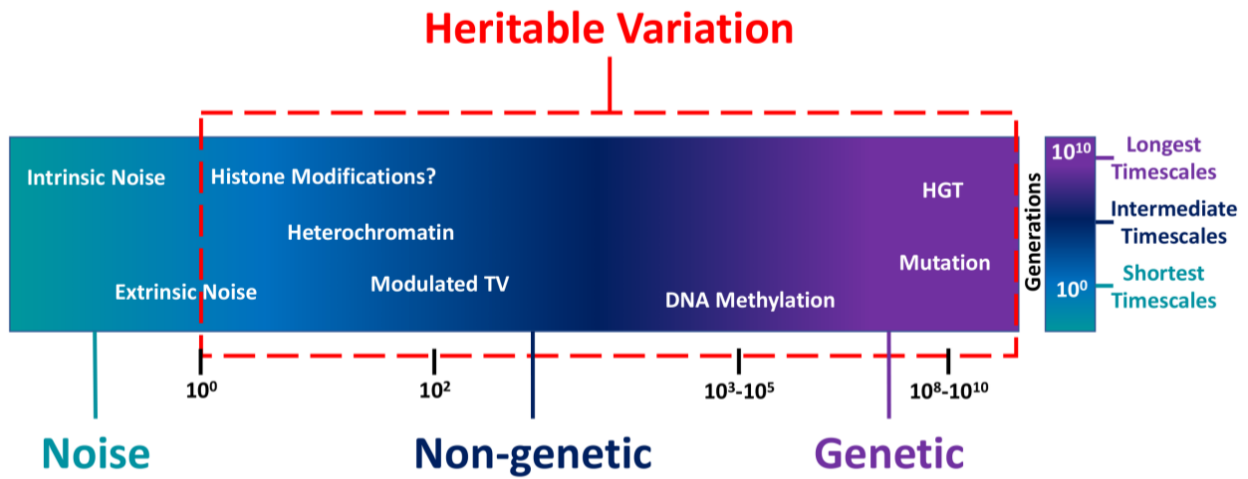

**Figure 1.** Timescales and heritability of antimicrobial drug resistance mechanisms. Drug resistance mechanisms are categorized into unmodulated gene expression "noise", "non-genetic" modulated transcriptional variability (TV) and epigenetic mechanisms, and "genetic" mechanisms. The approximate timescale of inheritance for each of these mechanisms is shown in units of cellular generation. The question mark signifies that histone modifications may not be heritable on their own and may require other epigenetic processes for heritability [42]. HGT is an acronym for horizontal gene transfer.

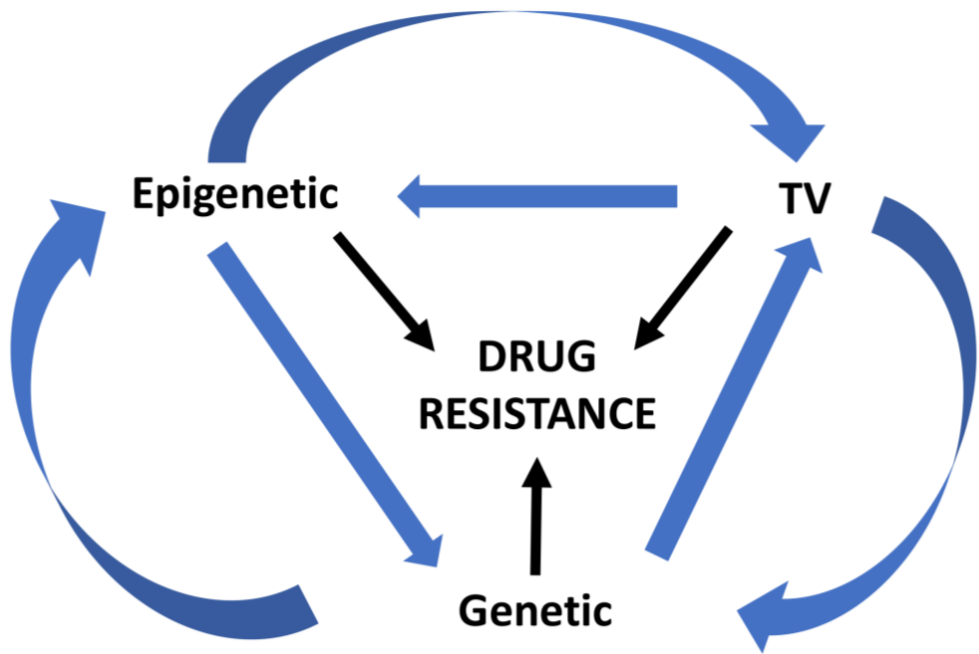

**Figure 2.** Proposed interplay between different forms of antimicrobial drug resistance. Epigenetic, modulated transcriptional variability (TV), and genetic mechanisms are all known to affect drug resistance (black arrows). The blue arrows denote an interaction between these different forms of drug resistance, which are each supported by varying degrees of evidence in the literature (see main text for details).